\documentclass[aps,twocolumn]{revtex4}
\usepackage{graphicx}
\begin{document}
\title{Monte Carlo Renormalization Group for Entanglement Percolation} 
\author{ Duygu Balcan$^{1}$ and Ay\c se Erzan$^{1,2}$}
\affiliation{$^1$  Department of Physics, Faculty of  Sciences
and
Letters\\
Istanbul Technical University, Maslak 80626, Istanbul, Turkey }
\affiliation{$^2$  G\"ursey Institute, P. O. Box 6, \c
Cengelk\"oy 81220, Istanbul, Turkey}
\date{\today}

\begin{abstract}

We use a large cell Monte Carlo Renormalization
procedure to compute the critical exponents of a system of growing linear
polymers.  We simulate the growth of non-intersecting chains in large MC
cells. Dense regions where chains get in each others' way, give rise to
connected clusters under coarse graining. At each time step, the fraction
of occupied bonds is determined in both the original and the coarse
grained configurations, and averaged over many realizations. Our results
for the fractal dimension on three dimensional lattices are
consistent with the percolation value.

\end{abstract}

\maketitle

\section{Introduction}

The gelation process in crosslinked polymers has been studied within the
context of percolation
theory~\cite{Stauffer1,Stauffer2,Herrmann1,Herrmann2}, but the
rheokinetics~\cite{rheokinetics} of bulk linear polymers as a function of
chain length has recieved less theoretical attention. As the density of
chains, and their length increases, the viscosity starts to increase well
before the onset of vitrification.~\cite{Cicerone} It has been
proposed~\cite{deGennes,Doi} that this is due to the entanglement of the
polymer chains.

We would like to pose the question of whether the percolation of
entanglement clusters is in the same universality class as percolation. We
define entanglement clusters starting from ordinary sets of connected
bonds. We will consider two such sets connected, as long as they come with
a lattice constant of each other, i.e., share the end points of an empty
bond. Since the length of the chains, or the average length between
entanglements, could introduce a second length scale into the problem,
this could potentially lead to a crossover to a different universality
class than percolation.

It has
been found in two dimensions~\cite{ConiglioDaoud79} that the vulcanization
process, which involves the crosslinking of long chains, 
 is in the percolation universality class,
and that there is a crossover between Self Avoiding
Walk and percolation behaviour as a function of the fugacity of the
crosslinkers. Similarly, Jan et al.~\cite{Jan} find in three 
dimensions that the  crossover from SAW to percolation exponents 
already occurs for any finite value of the concentration of initiators 
(from which the chainlike structures grow), for a mixture of monomers with 
functionalities $\ge 2$, i.e., again in the presence of crosslinkers.
A Monte Carlo simulation in two dimensions reveals~\cite{Bunde} that a 
growth model can cross over from SAW like behaviour to percolation, as a 
function of the cluster mass.

The percolation of clusters which are not necessarily
connected but linked to each other by loops, has also been studied by very
large MC computations~\cite{Kantor}. These authors find that the new
critical point is very close to the ordinary percolation threshold on the
cubic lattice, with the eigenvalue of the
renormalized occupation probability being indistinguishable from that of 
the ordinary percolation problem.
   
In this paper we introduce a special Monte Carlo (MC) Renormalization
Group procedure to investigate the universality class for the entanglement
phase transition of a linear polymer system, in three dimensions and with 
no crosslinkers (or monomers with functionality greater than 2) present.  
The polymerization process
is modeled by growing non-intersecting chains from a set of randomly
chosen sites on a cubic lattice. Chains which occupy nearest neighbor
sites on the lattice are considered to be part of the same connected
cluster, and coarse grain to occupied bonds.

In the next section, we present the simulations and the MC renormalization
group procedure.  In section 3 we present an analysis of the results.  We
conclude with a discussion in section 4.

\section{MC simulations and the Renormalization Group}

In this section we describe a Monte Carlo renormalization group procedure, 
to deal with the 
percolation of entangled clusters of linear polymers.
Since the long linear chains of the growing bulk polymer cannot be accomodated in small cells, 
we start by simulating the polymer growth on relatively large lattices. 
However we make a different choice than the one made by 
Swendsen~\cite{Swendsen1,Swendsen2,Swendsen3} in the way that the MC 
renormalization group is introduced, as illustrated schematically in Fig.1.

\begin{figure}
\leavevmode
\rotatebox{0}{\scalebox{1.0}{\includegraphics{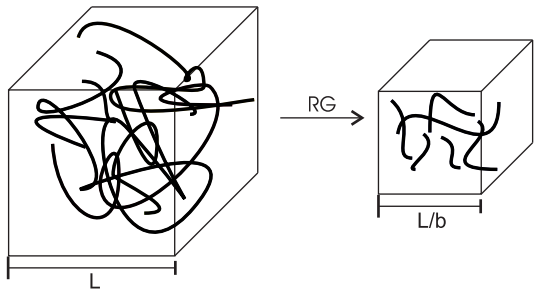}}}
\caption{We use a set of kinetic growth rules, to grow a mass of 
non-intersecting linear chains in 3 dimensions. Then we apply an 
RG transformation to get the new coarse-grained system.} 
\end{figure}

The MC Renormalization Group procedure which we use consists of paving the 
large MC cell by 
small cells on which one performs a  coarse graining 
transformation.~\cite{Reynolds1} This  induces an RG 
transformation on the bond occupation probability, averaged over many independent 
realizations. 

In the conventional MC renormalization group procedure,
the coarse graining is done in one
step~\cite{Swendsen1,Swendsen2,Swendsen3}. 
In the case of percolation, this would mean coarse graining 
the large MC cell to one 
occupied or empty bond - i.e., one
asks the question whether the cell is spanned by a connected cluster or
not.  Adopting the conventional definition of a connected cluster, this
one-step coarse graining procedure is clearly not appropriate in our case,
since for non-intersecting linear chains, the MC cell will be spanned in
case there is at least one chain that grows across it, and otherwise not.
This does not describe the physical problem at hand.

The effect which we are trying to model has to do with linear chains that
do not intersect, but nevertheless get physically close to each other in
dense regions where they constrain each others' motion, at least on short
time scales, and thus form an effective three-dimensional network.  We
look, therefore, for a way in which we can capture this effectively three
dimensional behaviour, and the answer lies in making a coarse graining 
with a scale factor $b$ smaller than the linear size $L$ of the large 
cell.
We will illustrate below, that chains which are non intersecting can,
under coarse graining, go to branched structures, which behave very much
like ordinary percolation clusters.

\subsection{The Kinetic Growth Model for  Linear Polymers}

We have generated our configurations on a cubic 
lattice with linear dimension $L$ and periodic boundary conditions. 
We start from an empty lattice, and then place $C$ initiators at randomly chosen lattice sites. As the chains grow, the initiators are displaced in the growth direction and mark the ``active end" of the chains. We don't allow a chain to 
intersect itself or the other chains during this process.  If the chain hits a dead end, it stops growing.

The growth rules are as follows. During one 
time step, from time $t$ to $t+1$, we consider each chain one by one, 
look at the active end of the chain and count the number of empty 
nearest neighbor sites. If this number is zero, the chain is trapped, 
and we proceed to the next chain. If the number is greater than zero, 
with {\em growth probability} $k$ we extend the chain to one of these empty sites, 
chosen at random. (This extra parameter $k$ gives us the possibility of probing arbitrarily small increments in the occupation probability on our finite cell.)We mark this new location as the active end of the 
chain, and  place a bond between the old and the new location of this 
active end. After this process has been applied to all the chains, one gets the configuration at time $t+1$. 

We use this set of kinetic growth rules, to grow a mass 
of non-intersecting linear chains in relatively large cells, consisting of cubic lattices of linear size $L=8,16,32$.

\begin{figure}
\leavevmode
\rotatebox{0}{\scalebox{1.0}{\includegraphics{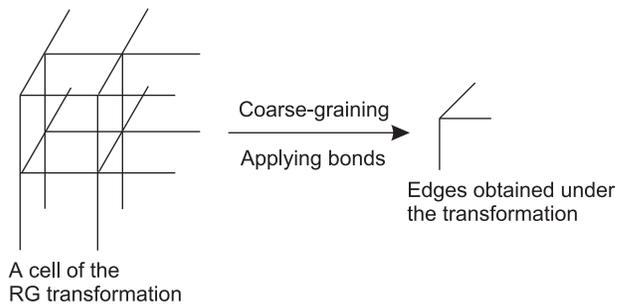}}}
\caption{One cell of the RG transformation. Under our coarse-graining rule 
one cell goes to at most three bonds in the three Cartesian directions.}
\end{figure}

\subsection{Compactification and Coarse-Graining}

In Fig. 2, we display the small cells with which our lattice will be paved, in order to perform the coarse graining transformation.
This choice for the coarse graining rule has been used by Reynolds et 
al.~\cite{Reynolds2} and Bernasconi~\cite{Bernasconi}. Under our coarse-graining 
rule each cell (with $3 b^3$ bonds, where $b$ is the rescaling factor, here chosen to be 2) goes to at most three bonds 
in the coarse-grained lattice. 

Before we coarse grain, we perform a compactification of 
our randomly connected clusters by connecting nearest neighbor occupied 
sites, which are not already connected by a bond.(See Fig. 3) This is motivated by the fact that in dense regions, chains which physically get in each others' way are not chemically connected, but nevertheless contribute to the instantaneous shear modulus of the effective network.

\begin{figure}
\leavevmode
\rotatebox{0}{\scalebox{1.0}{\includegraphics{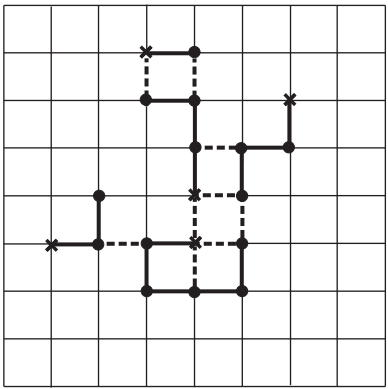}}}
\caption{An illustration of our linear polymer system in 2 dimensions 
at time t. Solid lines show the 
occupied bonds, solid dots indicate the monomers. Cross marks show the active 
ends of the chains, from which a chain grows. The  dashed 
lines show the bonds added in the compactification procedure; these are included in the computation of  $p(t)$.} 
\end{figure}

We determine the state of the bonds in 
the coarse-grained lattice by considering the configurations in each 
small cell, as illustrated in Fig. 4. We look for a path which spans the 
cell in each Cartesian direction. If there is at least one such  path this 
cell 
goes to an occupied bond in this direction in the coarse-grained 
lattice. Otherwise, it goes to an empty bond. By applying this rule to 
each small cell in the original lattice we get the new coarse-grained 
lattice. 

\begin{figure}
\leavevmode
\rotatebox{0}{\scalebox{1.0}{\includegraphics{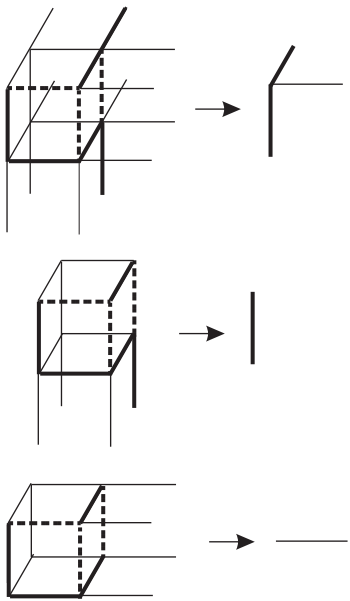}}}
\caption{An example of our coarse-graining rule. In each Cartesian 
direction, we look for a path which spans the cell in this direction. See text.} 

\end{figure}

It can be noticed in Fig. 4 that the occupied bonds spanning the small cell in the horizontal and the vertical directions are not chemically connected, and yet, (even without the bonds added for compactification) they go, under coarse graing to an occupied junction.

We now compute  the fraction of occupied bonds 
in both the original and coarse-grained lattices, $p(t)$ and 
$p'(t)$, where we also count the bonds added in the compactification process.  
These are averaged  over many independent runs, to find 
$<p(t)>$ and $<p'(t)>$, for each time $t$.  Eliminating $t$, the coarse graining procedure leads to a renormalization group transformation on the average occupation probability, as shown in Figure 5.
The fixed point of the RG transformation and the RG eigenvalue are 
computed from the MC  data, leading to the value of the critical exponent we are interested in.

\section{Finite size scaling and the fractal dimension}

The 
correlation length, 
\begin{equation}
\xi \sim{(p-p^\ast)}^{-\nu}\;\;\;, \label{xi}
\end{equation}
for $p$ very close to the critical point $p^\ast$ (the fixed point of the
RG transformation) exceeds $L$. From then on, $\xi \sim L$, i.e., it
behaves like a constant with respect to $(p-p^\ast)$. In other words, the
relationship in Eq.(\ref{xi}) breaks down and a finite size scaling
analysis is in order.~\cite{Stauffer1}

The mass contained in the incipient infinite cluster, $\Delta M \sim 
L^{D_f}$, where $D_f$ is the fractal dimension of the percolation 
cluster,so that $P_\infty = \Delta M / V \sim L^{-\beta / \nu} \sim L^{ 
D_f - d }$. We  may also show that the mass $M_0$ contained in the rest of 
the finite clusters scales like $M_0/V \sim L^{(2-\tau)D_f}$, with 
$(2-\tau)D_f=D_f-d$. Thus, the total concentration $p= (\Delta M + M_0)/V 
\equiv M/V$ of occupied bonds 
scales like 
\begin{equation}
p\sim L^{D_f-d}\;\;\;. \label{fractal}
\end{equation}
 We will make use of this fact to compute the fractal dimension of the
percolation cluster, which can be directly related to the other known
critical exponents, e.g.  via $D_f=(\beta+\gamma)/\nu$.~\cite{Herrmann2}

From (\ref{fractal}), we see that under a
rescaling transformation $L \to L'$=$L/b$, $M \to M' \sim b^{-D_f} M$, so
that
 \begin{equation}
  p'=\frac{M'}{V'} \sim b^{d-D_f}p\;\;\;.
\end{equation}
At the fixed point of the RG transformation, $p^\ast$, with $\lambda\equiv
dp^\prime / dp \vert_{p^\ast}$, one has, $\lambda=b^{d-D_f}$, or
\begin{equation} 
D_f=d-{\frac{\ln \lambda}{\ln b}}\;\;\;. \label{Df}
\end{equation} 
In Fig. 5, we plot $<p'(t)>$ versus $<p(t)>$ for $L$=32
where the averages have been performed over $10^4$ independent runs, with
$C=0.02 \times L^3$ and $k=0.01$.  Notice that $p^\ast \approx 0.02$ is
much smaller than the expected value for percolation on a cubic lattice
(0.2488)~\cite{Stauffer1}.  We have varied $C$ between 1 to 9$\%$ of the
lattice points, and a fixed point $p^\ast$ has been found for all these
different $C$ values, although the value of $p^\ast$ descreases with
decreasing $C$. This can be seen to result from an effective length scale
introduced by the increasingly long chains growing from fewer and fewer
initiators, i.e., an effective lattice spacing of the order of
$C^{-1/3}/L$~\cite{Jan}.

\begin{figure}
\leavevmode
\rotatebox{0}{\scalebox{0.8}{\includegraphics{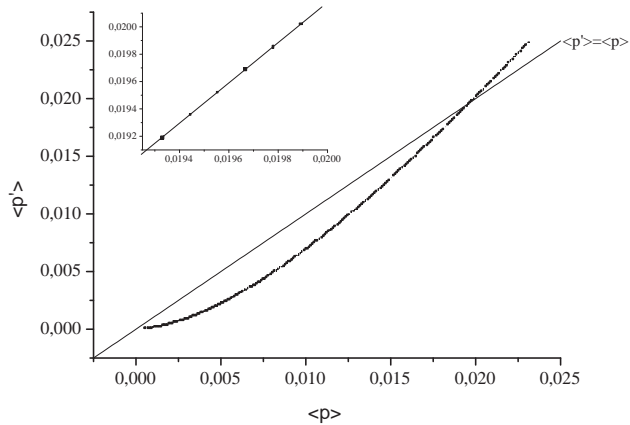}}}
\caption{A plot of $<p'>$ v.s. $<p>$ for the 2$\%$ concentration of initiators on an $L=32$ lattice.  The data points are averaged over $10^{4}$ independent runs. The inset shows the same graph in the vicinity of the fixed point $p^\ast$.}
\end{figure}

We obtain $\lambda$ by making a linear fit to the data around this point,
and we obtain the value of the critical exponent $D_f$ from Eq.
(\ref{Df}). Our results are given in Table 1. Within our error bars we
have not found any dependence of the RG eigenvalue on the concentration of
initiators, within the range of values we have considered.  The errors
reported in Table 1 are those arising from taking one standard deviation
of each of our data points reported in the inset of Fig. 5.  The best
value of $D_f$ for percolation clusters in three dimensions, obtained from
simulation results~\cite{Stanley} is $2.53\pm 0.03$. Given that we are
performing a finite cell real space renormalization group calculation, our
results are consistent with the percolation fractal dimension.

\begin{table} 
\begin{ruledtabular} 
\caption{$\lambda$ and $D_f$ values
obtained by our simulations.} \begin{tabular}{c|c|c|c} {\boldmath ~$L$~}&8
& 16 & 32 \\ \hline\hline $\lambda$ &1.476& 1.479 & 1.472 \\ $D_f$ &2.438&
2.436 & 2.442 \\ $\Delta$$D_f$ &0.004& 0.002 & 0.001 \\ 
\end{tabular}
\end{ruledtabular} \end{table}

\section{Discussion}

The physics of bulk linear polymers is an extremely interesting and
rapidly growing field. As the average molecular weight (or chain length)
grows, bulk linear polymers are known to exhibit many of the properties of
ordinary gels, such as resistance to shear and the capacity to take up
solvent and swell, while retaining their original shape.~\cite{Yilmaz}
This behaviour is present even in linear polymers like PMMA (Poly-methyl
methacrylate) where inter-chain interactions~\cite{Saiani,Erukhimovich}
are extremely weak, and therefore must arise from purely geometrical
effects, in other words entanglements.~\cite{Zhu,Tian} In this context we use ``entanglement" to mean a dense region where chains physically
impede the motion of other chains, and not in the mathematically precise
sense of knots~\cite{Kantor,Holroyd1,Holroyd2,Grimmet}.

The transition to a regime where the presence of an entanglement network
can be ascertained via dilatometric techniques~\cite{Okay1} takes place
before or around the onset of the ``gel effect,"  namely the characteristic
rapid growth in the rate of polymerization~\cite{Okay2,Okay3}. This effect
arises due to the transition to a diffusion-limited regime, where end to
end termination of chains is essentially suppressed, so that single chains
grow rapidly~\cite{Okay1,Okay2,Okay3,Zhu}.  Its onset has been conjectured 
to be due to the entanglement of the linear chains~\cite{deGennes,Doi}.

Defining entanglement clusters as chains which will be 
considered connected if they happen to pass through nearest neighbor 
sites, we set out to study the scaling behaviour in the vicinity of the 
entanglement percolation threshold in three dimensions.
We have demonstrated, via a modified MC Renormalization Group, that for
realistic concentrations of initiatiors,  entangled 
clusters
of linear chains have the same fractal dimension as ordinary percolation,
although the entanglement percolation threshold is  much lower than
for standard bond percolation on a cubic lattice.

{\bf Acknowledgements}

We would like to thank Nihat Berker for useful comments.  One of us would
like to acknowledge partial support from the Turkish Academy of Sciences.

\end {document}